\documentclass[aps,pre,twocolumn,superscriptaddress,showpacs]{revtex4}
\usepackage{graphicx}
\usepackage{amssymb}

\begin{document}
\bibliographystyle{apsrev}

\title{Capillary Adhesion at the Nanometer Scale}
\author{Shengfeng Cheng}
\email{chengsf@vt.edu}
\affiliation{Department of Physics and Astronomy, Johns Hopkins University,\\
3400 N. Charles Street, Baltimore, Maryland 21218, USA}
\affiliation{Department of Physics, Virginia Polytechnic Institute and State University, Blacksburg, Virginia 24061, USA}
\author{Mark O. Robbins}
\email{mr@pha.jhu.edu}
\affiliation{Department of Physics and Astronomy, Johns Hopkins University,\\
3400 N. Charles Street, Baltimore, Maryland 21218, USA}

\date{\today}

\begin{abstract}
Molecular dynamics simulations are used to study the capillary adhesion 
from a nonvolatile liquid meniscus between a spherical tip and a
flat substrate. 
The atomic structure of the tip, the tip radius, 
the contact angles of the liquid on the two surfaces, and the volume of the liquid bridge are varied. 
The capillary force between the tip and substrate is calculated 
as a function of their separation $h$. 
The force agrees with continuum predictions for $h$ down to $\sim 5$ to $10$nm. 
At smaller $h$, the force tends to be less attractive than predicted and has strong oscillations.
This oscillatory component of the capillary force is completely missed in 
the continuum theory, which only includes contributions from the surface tension around the circumference of the meniscus 
and the pressure difference over the cross section of the meniscus.
The oscillation is found to be due to molecular layering
of the liquid confined in the narrow gap between the tip and substrate. 
This effect is most pronounced for large tip radii and/or smooth surfaces. 
The other two components considered by the continuum theory are also identified.
The surface tension term, as well as the meniscus shape, 
is accurately described by the continuum prediction for $h$ down to $\sim 1$nm,
but the capillary pressure term is always more positive 
than the corresponding continuum result. 
This shift in the capillary pressure reduces the average adhesion by a factor as large as $2$ from its continuum value 
and is found to be due to an anisotropy in the pressure tensor. 
The cross-sectional component is consistent with the capillary pressure predicted by the continuum theory (i.e., the Young-Laplace equation), 
but the normal pressure that determines the capillary force is always more positive 
than the continuum counterpart.
\end{abstract}


\maketitle

\section{Introduction}

Adhesion and friction due to capillary forces 
affect the flow and yield properties 
of granular materials to a large extent \cite{jaeger92,jaeger96,bocquet98,herminghaus05}.
One example is that the strength of powders can be 
greatly enhanced by absorbing moisture from the ambient atmosphere \cite{rabinovich02}. 
This fact has been known to our ancient ancestors when they built walls from clay mixed
with an appropriate amount of water. The meniscus formed between clay particles 
binds them together with capillary forces, increasing 
the unconfined yield strength of the granular assembly. 
The same physics is involved when wet sand is used to build sandcastles, 
which would not be possible to make with dry sand 
\cite{hornbaker97,albert97,halsey98,nowak05,scheel08}.

As dimensions shrink into the nanoscale, capillary forces become increasingly important 
and can be the dominant source of adhesion between surfaces \cite{yang06,persson08,zwol08}. 
For example, they often prevent micro/nano-electromechanical systems from functioning
under ambient conditions or lead to damage in their fabrication processes \cite{charlaix10}. 
They play a major role in nanoscopic sliding friction and lead to
a nontrivial velocity dependence of friction \cite{binggeli94,ohnishi02,riedo02,riedo05, noel12}. 
Many previous experiments also show that 
when an atomic force microscope (AFM), or more generally any small probe, is used to examine a 
hydrophilic surface in a humid environment, 
a major contribution to the tip-surface interaction is 
from the capillary forces associated with the water meniscus bridging the tip and surface  
\cite{thundat93,binggeli94,eastman96,piner97,piner99,xiao00,jones02,sirghi03,feiler07,yangsh08}.
The geometry in these experiments can be reasonably approximated as
a liquid bridge connecting a spherical tip of certain size and a flat surface. 
Though there can be complex interplay 
between the tip shape and capillary forces \cite{thundat93,butt06}, 
the simple sphere-on-flat geometry 
serves as a prototype to understand more realistic situations. 
These include particle-surface 
or particle-particle interactions 
occurring in many physical, chemical, and biological processes \cite{israelachvili91}.

A capillary bridge can form either by capillary condensation or by accumulation
of liquid \cite{butt09}. The former occurs 
when multilayer adsorption from a vapor proceeds to the point
at which a small gap, such as the one between the tip and sample surfaces, 
is filled with liquid separated from the vapor phase by
menisci \cite{IUPAC}. 
Here the presence of the surfaces shifts the phase diagram of the fluid to favor the liquid phase \cite{charlaix10}.
Capillary condensation 
is responsible for the formation of water menisci in the AFM experiments mentioned above 
\cite{thundat93,binggeli94,eastman96,piner97,piner99,xiao00,jones02,sirghi03,feiler07,yangsh08}.
The capillary pressure inside the liquid bridge 
is fixed by the relative humidity of the vapor phase, 
and the volume of the liquid bridge depends on the size of the gap. 
Capillary condensation also occurs 
in liquid crystals \cite{kocevar01,stark04}, 
aqueous mixtures of polymers \cite{olsson04,sprakel07}, and 
model fluid mixtures \cite{bucior03} when confined in small gaps.

A capillary bridge can also form by accumulation of liquid.
For example, insects excrete fluid to create adhesion that allows them to 
walk on walls or ceilings. 
In this case and many other occasions, a nonvolatile liquid is used 
in order to avoid the quick dry-out of the liquid 
\cite{willett00,malotky01,rabinovich02,wang09}. 
When the liquid only partially wets the solid surfaces, 
the volume of the liquid bridge is controlled by 
the amount of liquid injected. 
The bridge volume and the wetting properties of the solid surfaces then
determine the shape of the liquid bridge and the capillary pressure. 
In this case the volume of the liquid bridge acts as a controlling parameter 
and determines the capillary forces \cite{scheel08,grof08}.

The continuum theory of capillarity, which is briefly introduced in the next section, 
is routinely used to calculate capillary forces 
and interpret experiments \cite{rowlinson89}. 
However, it is not obvious if the continuum theory is applicable 
to liquid bridges with nanometer dimensions. 
In this paper, we use MD simulations to study the
capillary forces produced by a small liquid bridge 
connecting a spherical tip and a flat substrate. 
Our goal is to address the fundamental limits of the continuum theory 
and to develop understanding of new behavior that enters at the nanometer scale.
We find that the interface shape and the interfacial contribution to the capillary force are surprisingly close to continuum theory in all cases.
However the total capillary force is strongly affected by molecular layering and pressure anisotropy when the separation between surfaces is less than a few nanometers.

The next section briefly describes the continuum theory.
Section III describes our simulation techniques and geometries.
Section IV presents results for a wide range of cases and the final section
presents a summary and conclusions.

\section{Continuum theory of capillary forces}
\label {continuum_theory}

\begin{figure}[htb]
\centering
\includegraphics[width=3in]{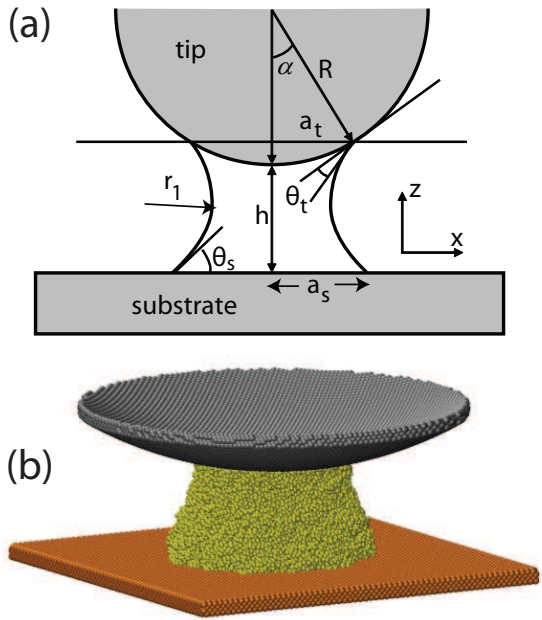}
\caption{(Color online) (a) Geometry of a liquid bridge between a
sphere and a substrate, including contact angles $\theta_i$ and radii $a_i$ 
at the intersection of the meniscus with the tip ($i=t$) and substrate ($i=s$).
The pressure drop is related to the radii of curvature through Eq. \ref{Laplace}.
The radii are negative when the center of the circle is outside the drop,
as for $r_1$ in the figure.
(b) A snapshot from simulations of a liquid bridge (yellow) 
between a spherical tip (gray) 
and an atomically flat substrate (orange). 
Here the tip was made by bending a thin fcc crystal with a (001) surface into a sphere
with radius $R=100\sigma$.
The bridge intersects the tip with radius $a_t \approx 25\sigma$.
Only the central parts of the substrate and tip are shown.
}
\label{AtomicConfig}
\end{figure}

The geometry of the system studied in this paper is sketched in Fig.~\ref{AtomicConfig}(a), 
where a liquid meniscus bridges between a spherical tip and a flat surface.
In the continuum theory, the pressure change $\Delta p$ across a curved liquid/vapor interface
(or more generally, a fluid/fluid interface)
satisfies the Young-Laplace equation:
\begin{equation}
\Delta p = \gamma (1/r_1 + 1/r_2) \equiv 2 \gamma \overline{\kappa}~,
\label{Laplace}
\end{equation}
where $\gamma$ is the liquid/vapor interfacial tension,
$r_i$ the principal radii of curvature, and $\overline{\kappa}$
the mean curvature.
We take $\Delta p$ to be negative when the interior of the drop has a lower
pressure. Then $r_i$ is negative when the center of the corresponding
circle is away from the drop as for $r_1$ in Fig. \ref{AtomicConfig}(a). 

For micrometer and smaller drops, pressure changes
due to gravity can be neglected.
The equilibrium interface must then have a constant mean
curvature so that the pressure is constant in both the liquid
and vapor phases.
The interface must also intersect solid walls at a contact
angle $\theta$ determined by the Young equation:
\begin{equation}
\gamma \cos\theta = \gamma_{sv}-\gamma_{sl}~,
\label{Young}
\end{equation}
where $\gamma_{sv}$ and $\gamma_{sl}$ are the solid/vapor and
solid/liquid interfacial tensions, respectively.
The Young equation represents a balance of tangential forces on the
contact line where all three phases meet
\cite{degennes85,thompson93b}.

The shape of the liquid bridge between a spherical tip
and a flat substrate is uniquely determined by $\Delta p$, the
tip-substrate separation $h$,
and the contact angles $\theta_{t}$ and $\theta_{s}$ of the liquid
on the tip and substrate surfaces, respectively.
Our MD results were compared to the exact continuum solution of Orr {\it et al}.
for Eq.~\ref{Laplace}
in terms of elliptic integrals \cite{orr75}.
Simpler analytic solutions are also widely used in analyzing experimental data
\cite{marmur93,gao97,lazzer99,stifter00,lambert08,israelachvili91},
but are only valid in limiting cases such as $|r_2| \gg |r_1|$
or constant $r_1$.

For nonvolatile liquids, $\Delta p$ is fixed by the total volume
$V_l$ of the liquid drop.
Volatile liquids can condense or evaporate until the chemical
potential of molecules in the gas and liquid phases is equal.
The value of $\Delta p$ then satisfies the Kelvin equation
that relates $\Delta p$ to the relative humidity of the vapor.
We will consider
the nonvolatile case because of the ease of equilibration.
The results for volatile liquids are the same for a given
interface geometry, but the variation of $\Delta p$
with $h$ is different in the two cases.

In the continuum picture there are two contributions to the capillary force $F_{cap}$
between the solids \cite{footm2}.
The first comes directly from the interfacial tension
exerted by the meniscus on the contact line where it contacts the solid.
This force is always attractive in the geometries considered,
and we will choose our sign convention so that an attractive force has a negative value.
The second contribution comes from the integral of the Laplace pressure over the
area where the drop contacts the solid.
This force is attractive when $\Delta p$ is negative.
Because of the azimuthal symmetry of the system about the vertical axis through the
sphere center, only the component of the force along this axis
is nonzero and
the interface intersects the tip and substrate at circles of
radius $a_t$ and $a_s$, respectively.
The sum of the meniscus and Laplace contributions gives:
\begin{equation}
F_{cap}= - 2\pi \gamma a_i \sin \phi_i + \pi a_i^2 \Delta p~,
\label{fcap}
\end{equation}
where $\phi_i$ is the angle relative to the $x$-$y$ plane
of the tangent to the liquid/vapor interface at the solid surface.
At the substrate $\phi_s = \theta_s$, while on the spherical tip of radius $R$,
$\phi_t = \theta_t + \alpha$, where $\alpha=\arcsin(a_t/R)$ (Fig. \ref{AtomicConfig} (a)).
Newton's third law requires that the magnitudes of the capillary
force on the tip and substrate are the same.
For a case like Fig. \ref{AtomicConfig} (b), the Laplace pressure acts
on a smaller projected area at the tip surface than at the substrate surface (i.e., $a_t<a_s$),
but this difference is compensated
for by the differences in $\phi_i$ and $a_i$.

Many macroscopic experiments are in the limit where the drop dimensions
are much smaller than the tip radius,
but much larger than the separation $h$.
In this limit ($R \gg |r_2| \gg |r_1|$), $F_{cap}$ is independent
of both $h$ and the drop volume $V_l$
\cite{israelachvili91}. 
As $V_l$ increases, the increases in $a_t$ and $a_s$ are compensated for
by the decrease of $\overline{\kappa}$ and thus $\Delta p$.
This condition frequently applies for granular materials held together
by liquid bridges formed via capillary condensation, making their mechanical properties relatively
insensitive to the relative humidity of the environment or the addition of extra nonvolatile liquids
\cite{eastman96,scheel08}.
However, our simulations are generally in a different limit where the drop and tip
have comparable dimensions and $F_{cap}$ depends on both $h$ and $V_l$.

Previous work has tested some
aspects of the continuum theory down to atomic scales.
For example, the Kelvin equation is obeyed by cyclohexane or water menisci
with radii of curvature as small as
$4$nm \cite{fisher79,fisher81a,fisher81b,kohonen00,mitropoulos08}.
MD simulations confirmed that the Young-Laplace and Young equations
hold down to a similar length scale for simple model potentials \cite{thompson93b,bresme98,takahashi13}.
Grand canonical Monte Carlo simulations have also been used
to investigate capillary condensation
and to calculate capillary forces \cite{stroud01,jang03}. 
However, the latter studies used lattice models on square or cubic grids
that may not capture the nanoscale behavior of atoms
and molecules.
To the best of our knowledge there are no molecular dynamics (MD) simulations reported on capillary condensation. 
One reason is that it is extremely difficult to study the thermal equilibration
of a liquid-vapor system with MD.

\section{Simulation Methods}

Figure~\ref{AtomicConfig}(b) shows a snapshot of
a liquid bridge that connects a spherical tip and a flat substrate. 
The liquid partially wets both the tip and substrate surfaces. 
In our simulations, the gap width $h$ between the tip and substrate is controlled
and the bridge-induced capillary force $F_{cap}$ between the two
is calculated as $h$ is varied. 

Since our goal is to address the generic behavior, we use a simple
model potential that captures many aspects
of the behavior of hydrocarbon
chains \cite{kremer90,thompson92,deconinck95,rottler01,cheng12jcp,ge13}.
Fluid molecules are modeled as short linear chains of 4 spherical beads.
All beads, except those pairs directly bonded in one molecule, 
interact with a Lennard-Jones (LJ) potential 
\begin{equation}
\label{LJPotential}
V_{LJ}(r)=4\epsilon[(\sigma/r)^{12}-(\sigma/r)^{6}-(\sigma/r_c)^{12}+(\sigma/r_c)^{6}]~,
\end{equation}
where $r$ is the distance between beads.
The LJ potential is truncated and shifted at a cutoff radius $r_c = 2.2 \sigma$.
The binding energy $\epsilon$, diameter $\sigma$, and mass $m$ of the beads
are used to define all dimensions.
The neighboring beads in each chain interact through a purely
repulsive LJ potential with $r_c=2^{1/6}\sigma$ and
an attractive finitely-extensible nonlinear elastic (FENE) potential \cite{kremer90}
\begin{equation}
\label{FENEPotential}
V_{FENE}(r)=-\frac{1}{2} K R_0^2 {\rm ln}[1-(r/R_0)^2]
\end{equation}
with the canonical values of $R_0=1.5\sigma$ and $K=30\epsilon/\sigma^2$.
Monatomic LJ systems have a very high vapor pressure and 
are not well-suited for studies of liquid/gas interfaces \cite{cheng11jcp}.
Increasing the chain length to 4 beads leads to a negligible vapor pressure
without significantly increasing the molecular size or slowing down the dynamics.
We have verified that increasing the chain length to 8 beads does not change any of
the conclusions reached below.

To provide a rough mapping of our results to experiments
we use the facts that a typical hydrocarbon has
surface tension $\gamma \sim 25{\rm mN/m}$
and the separation between molecules corresponds to a
chain diameter $\sigma \sim 0.5{\rm nm}$.
In our simulations the surface tension is calculated with the Kirkwood-Buff expression
for a flat liquid/gas interface \cite{kirkwood49}.
The fluid is kept at a constant temperature of $T=0.7 \epsilon/k_{\rm B}$
that corresponds to the melting temperature of a pure LJ system \cite{allen87} and
is about twice the glass transition temperature of FENE chains \cite{rottler03c}.
The calculated value of surface tension is $\gamma = 0.88\epsilon/\sigma^2$, which
implies $\epsilon \sim 7\times 10^{-21}{\rm J} \simeq 0.044{\rm eV}$.
The temperature also maps to a reasonable value of $\sim 360K$.
Since the bead diameter of 0.5nm corresponds to three to four carbons
along the backbone \cite{kremer90},
$m \sim 10^{-25}$kg.
The characteristic LJ time, $\tau= \sqrt{m\sigma^2/\epsilon}$,
is then of the order of $2$ps.
The unit of force is $\epsilon/\sigma \sim 14$pN and the unit of pressure
$\epsilon/\sigma^3 \sim 56$MPa.
All the mappings of LJ and real units are summarized in Table.~\ref{table_map}.

\begin{table}
\centering
\begin{tabular}{|c|c|c|} \hline
physical quantity & LJ unit & SI value\\ \hline
energy & $\epsilon$ & $7\times 10^{-21}{\rm J}$ \\ \hline
length & $\sigma$ & $0.5\times 10^{-9}$m \\ \hline
mass & $m$ & $1\times 10^{-25}$kg \\ \hline
time & $\tau$ & $2\times 10^{-12}$s \\ \hline
force & $\epsilon/\sigma$ & $1.4\times 10^{-11}$N \\ \hline
surface tension & $\epsilon/\sigma^2$ & $2.8\times 10^{-2}$N/m \\ \hline
pressure & $\epsilon/\sigma^3$ & $5.6\times 10^{7}{\rm N/m^2}$ \\ \hline
temperature & $\epsilon/k_{\rm B}$ & $5.1\times 10^{2}$K \\ \hline
\end{tabular}
\caption{Rough mapping between LJ and real units}
\label{table_map}
\end{table}

The substrate is treated as either an elastic or rigid solid.
Atoms in the substrate form an fcc lattice with a (001) surface.
The nearest-neighbor spacing is $d=2^{1/6}\sigma$.
For the elastic substrate,
nearest neighbors interact through a harmonic potential 
\begin{equation}
\label{HarmPotential}
V(r)=\frac{1}{2}k(r-d)^2
\end{equation}
with $k=228\epsilon/\sigma^2$. 
This substrate corresponds to a solid with LJ interactions at the strength of 4$\epsilon$
and a Young's modulus of about 240 $\epsilon/\sigma^3$ \cite{cheng10pre,luan06b}.
Given the mapping above, this modulus is about 14GPa,
which means that our elastic substrate is harder than typical
hydrocarbon solids, but much softer than the tip-substrate pairs in most
AFM experiments.
The elastic substrate has dimensions $190.49\sigma \times 190.49\sigma \times 7.94\sigma$
with periodic boundary conditions imposed in the $x$-$y$ plane.
The depth has little effect on the results, for reasons discussed in the next section.
For the rigid substrate, the thickness can be reduced
to include only those atoms that are within the interaction range
of the fluid.

The tip is modeled as a rigid sphere.
The radius $R$ is varied from values $25 \sigma \sim 13$nm, corresponding to a
sharp AFM tip, up to infinity, corresponding
to a flat surface.
The atomic scale structure of the tip is also varied, since this
profoundly affects the contact between the tip and substrate,
either bare or with an adsorbed layer \cite{luan05,luan06a,cheng10pre}.
Some tips are made by bending
three (001) planes of an fcc crystal with nearest-neighbor spacing $d^\prime$
into a spherical shape.
The case $d^\prime = d$ will be referred to as a commensurate tip,
since its period matches that of the substrate.
A denser tip with $d^\prime = 0.94437 d$ is referred to as incommensurate.
We also make tips by cutting a spherical shell out of an fcc crystal
that has the same structure as the substrate,
or an amorphous solid with a density $1.0 m/\sigma^{3}$. 
The former is called a stepped tip since it has a terraced surface,
and the latter is called an amorphous tip.
In all cases, the thickness of the shell is reduced to include only atoms
that are within the interaction range of the fluid.
A purely repulsive LJ potential (i.e., $r_c=2^{1/6}\sigma$) is used for the direct interaction between
the tip and substrate, but this interaction never contributes to the total force in the
cases shown below since the separation between the tip and substrate atoms is always larger than $r_c$.

\begin{figure}[htb]
\centering
\includegraphics[width=3in]{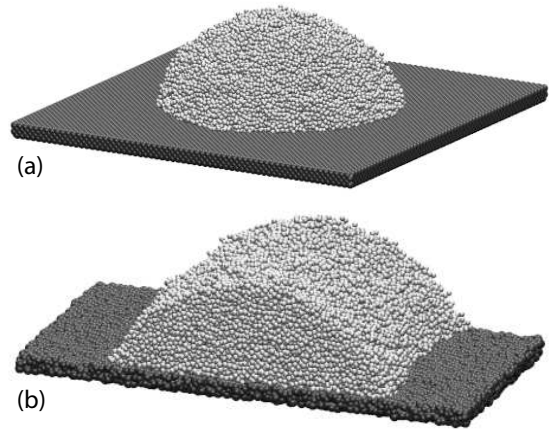}
\caption{Geometries used to calculate the contact angle $\theta$.
(a) A spherical drop (light) on a rigid crystalline substrate (dark).
The height of the drop is $\sim 23\sigma$ and
the radius of its base circle is $\sim 31\sigma$.
(b) A liquid cylinder (light) on a rigid amorphous substrate (dark).
The height of the cylinder is $\sim 21\sigma$, 
the length $47.6\sigma$, and the width
at its base $\sim 60\sigma$.}
\label{DropCylinder}
\end{figure}

The equilibrium contact angle $\theta$ is controlled by changing the interactions between
the liquid and solid.
Fluid beads and solid atoms interact with a LJ potential (Eq.~\ref{LJPotential})
with modified energy and length scales, $\epsilon_{fs}$ and $\sigma_{fs}$, respectively.
Unless noted, the simulations presented below use $\epsilon_{fs}=0.8 \epsilon$ and
$\sigma_{fs}=1.2\sigma$, and the potential is truncated and shifted at $r_c=1.8\sigma_{fs}=2.16\sigma$.
As discussed next, this combination of $\epsilon_{fs}$ and $\sigma_{fs}$ leads to a contact angle $\theta \sim 75^\circ$
for our crystalline substrate.
Increasing $\epsilon_{fs}$ favors wetting of the solid and decreases $\theta$.
Where noted, we also show results for commensurate tips with
$\epsilon_{fs}=1.08\epsilon$, which yields $\theta \sim 12^\circ$ for our substrate.
A further increase in $\epsilon_{fs}$ to $1.2\epsilon$ leads to complete wetting
(i.e., $\theta \rightarrow 0$) and the liquid film spreads over the entire substrate.

\begin{table}
\centering
\begin{tabular}{|c|c|c|c|} \hline
solids & $\epsilon_{fs}/\epsilon$ & contact angle ($^\circ$)\\ \hline
substrate or commensurate tip & 0.8 & 75 \\ \hline
substrate or commensurate tip & 1.08 & 12 \\ \hline
stepped tip & 0.8 & 75 \\ \hline
amorphous tip & 0.8 & 66.3 \\ \hline
incommensurate tip & 0.8 & 61.6 \\ \hline
\end{tabular}
\caption{Contact angles for solid geometries and fluid-solid interaction energies used in simulations.}
\label{table_angle}
\end{table}

\begin{figure}[htb]
\centering
\includegraphics[width=3in]{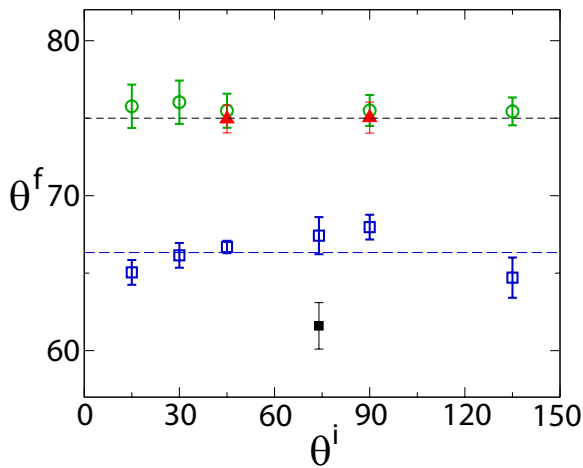}
\caption{(Color online) The final contact angle $\theta^f$ as a function
of the initial angle $\theta^i$ for different solid surfaces and liquid geometries.
For the fcc crystal with the nearest neighbor spacing $d=2^{1/6}\sigma$
(i.e., for the substrate, the commensurate tip, and the stepped tip),
results are shown for both a liquid
cylinder (circle) or a circular drop (filled triangle).
Cylinder results are shown for the amorphous solid (open square) and
the fcc crystal with the nearest neighbor spacing $d^\prime = 0.94437 d$ (filled square),
i.e. for the amorphous and incommensurate tips, respectively.
}
\label{ConAngPlot}
\end{figure}

Figure~\ref{DropCylinder} illustrates the geometries used to calculate $\theta$.
A circular drop or cylinder of liquid is placed on a flat substrate with
the desired structure and interactions.
Contact angles are obtained by fitting the liquid-vapor interface to a
sphere or cylinder and finding the angle at the solid/liquid interface.
The initial drop or cylinder makes an angle $\theta^i$ with the solid.
The relaxation of the drop or cylinder is followed 
and the final angle $\theta^f$ is calculated and taken as $\theta$. 
As shown in Fig.~\ref{ConAngPlot},
$\theta^f$ is independent of $\theta^i$.
At the macroscopic scale, disorder and other effects usually lead
to a dependence of $\theta^f$ on initial conditions known as
contact angle hysteresis \cite{degennes85}. 
Any hysteresis in our simulations is less than the errorbars ($\sim 1-2^\circ$) in $\theta^f$ .

Some continuum treatments include an additional energy equal to the
length of the contact line times a line tension \cite{churaev82,gaydos87}.
A positive line tension would favor a decrease in the circumference of 
a circular drop, leading to a larger contact angle than for a cylinder.
The calculated $\theta$ is the same for both geometries, implying that
the line tension has negligible effect.
Line tension may become important as the drop radius decreases to a 
few $\sigma$, but we will focus on capillary bridges that are at least as
wide as the drops used to find $\theta$.

The contact angles for tips with different atomic structures are calculated using
flat surfaces with the corresponding structure (Table \ref{table_angle}).
In principle the contact angle may change when a surface is bent into a sphere,
but tests indicated that any change is within the errorbars in $\theta$ for
the tip radii $R \ge 25\sigma$ used here. 
In Fig.~\ref{ConAngPlot}, $\theta = 75^\circ$ for the substrate and the
commensurate tip.
We take the same value for the stepped tip, though the contact angle may
be different on terraces and terrace edges.
Atoms in the incommensurate tip have the same LJ potential with fluid beads 
as those in the commensurate tip, but the incommensurate tip has a slightly higher density,
which leads to stronger adhesion and a slightly lower angle $\theta=61.6^\circ$.
The amorphous surface has an intermediate angle of $66.3^\circ$.

The zero of tip-substrate separation $h$ is somewhat ambiguous at atomic
scales.
The separation between the centers of the closest atoms on the tip and substrate will be defined as $h_0$.
The space available to liquid atoms is smaller because of the repulsive excluded volume interactions from each solid wall.
We determine the width $h_{ex}$ of the excluded layer near each wall
so that the volume available to liquid atoms is consistent with the continuum
expression for $h=h_0-2h_{ex}$.
First the liquid density is calculated as a function of distance from
the wall.
There are strong oscillations near the walls due to the liquid layering
described below and in previous work \cite{horn81,thompson90a,toxvaerd81,magda85}.
These oscillations decay quickly to a constant density
$\rho_0 \sim 0.905m/\sigma^3$.
The value of $h_{ex}$ is calculated by requiring that the integral over the
oscillatory density from the wall to some central height
$z$ is equal to $\rho_0 |z-h_{ex}|$.
As long as $z$ is in a region with constant density $\rho_0$, it does not affect
the value of $h_{ex}$.
For the substrate and commensurate tip, $h_{ex} = 0.775\sigma$ at zero pressure.
Changes with pressure and tip geometry (incommensurate and amorphous) are
too small ($< 0.1 \sigma$) to affect the figures below.
Therefore we take $h_{ex} = 0.775\sigma$ for all the solid walls below.

All simulations were performed using the Large-scale 
Atomic$/$Molecular Massively Parallel Simulator (LAMMPS) 
developed at Sandia National Laboratories \cite{plimpton95,lammps}.
Interatomic forces are calculated with the help of neighbor lists.
A velocity-Verlet algorithm with a time step $dt=0.005\tau$ is used to integrate the equations of motion.
Constant temperature ($T=0.7\epsilon/k_{\rm B}$) is maintained with a Langevin thermostat 
with damping rate $\Gamma = 0.1\tau^{-1}$.

Since the tip/substrate interaction is truncated at $r_c=2^{1/6}\sigma$, there is no direct van der Waals attraction between solids.
This attraction would be $-AR/6h^2$ where $R$ is the tip radius and $A$
a Hamaker constant \cite{israelachvili91}.
Estimates of this force are smaller than the capillary forces 
in the examples shown below and decrease much more rapidly as $h$ increases.

\begin{figure}[htb]
\centering
\includegraphics[width=3in]{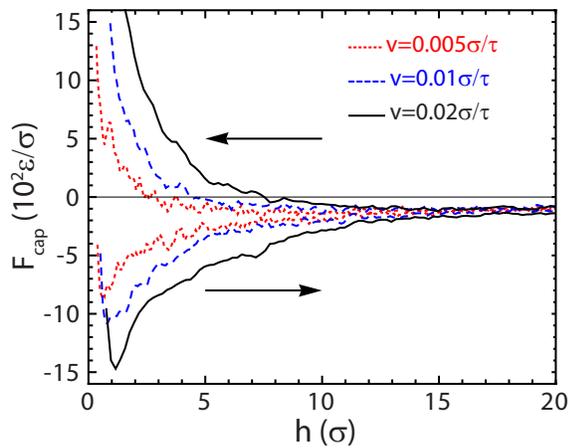}
\caption{(Color online) Force-separation curves for the amorphous tip
with $R=100\sigma$
show hysteresis as the tip is displaced at a constant velocity $v$ (legend)
toward ($h$ decreases as indicated by the left arrow) 
and away from ($h$ increases as indicated by the right arrow) the substrate.
}
\label{HysteresisPlot}
\end{figure}

In our first simulations the capillary force $F_{cap}$ is calculated as a function
of $h$ while the tip is brought toward and then away from the substrate at
a constant velocity $v$.
As shown in Fig.~\ref{HysteresisPlot}, the resulting force strongly
depends on the rate and direction of motion.
The hysteresis between approach and retraction increases as $h$
decreases and $v$ increases.
For most cases, it is even larger than the equilibrium force that
we wish to calculate.
Similar results are observed for other tips and liquid bridges.

One source of the hysteresis is the time required for fluid flow to equalize
the pressure in the drop.
From continuum lubrication theory, the force required to displace a viscous
fluid between a sphere and substrate scales as
$F_{visc} \sim 6 \pi R^2 \eta v /h$
where $\eta$ is the viscosity \cite{batchelor67,cai07}.
Our results are qualitatively consistent with this scaling,
although deviations grow at small $h$ where confinement may change
the viscosity \cite{gee90,thompson92,granick91}.
The force is also affected by the finite volume of the drop
and the dynamics of the contact line
on the two solid surfaces \cite{borkar91}.

To resolve forces with the desired accuracy would require decreasing
the velocity by two orders of magnitude or more relative to the lowest
$v$ in Fig. \ref{HysteresisPlot}.
While such velocities ($v = 5 \times 10^{-5}\sigma/\tau \sim 1$cm/s)
are still orders of magnitude faster than normal velocities
in typical AFM experiments, they are very hard to achieve with MD simulations.
We find it more efficient to decrease $h$ in small steps
and allow the fluid to relax before obtaining statistical averages
of local and global forces.
The required relaxation time increases as $h$ decreases and /or $R$ increases.
We find negligible hysteresis with an equilibration time of $1000 \tau$
for $h \gtrsim  1.5 \sigma$.
At smaller separations the fluid enteres a glassy state and the results are not reversible on any accessible time scales.
Similar glass transitions have been observed in Surface Force Apparatus (SFA)
experiments \cite{gee90,klein95,demirel96a}
and simulations \cite{thompson92,he99,gallo00,cheng10pre,alba06}.

\section{Results and Discussion}
\subsection{Effects of Substrate Elasticity}

\begin{figure}[htb]
\centering
\includegraphics[width=3in]{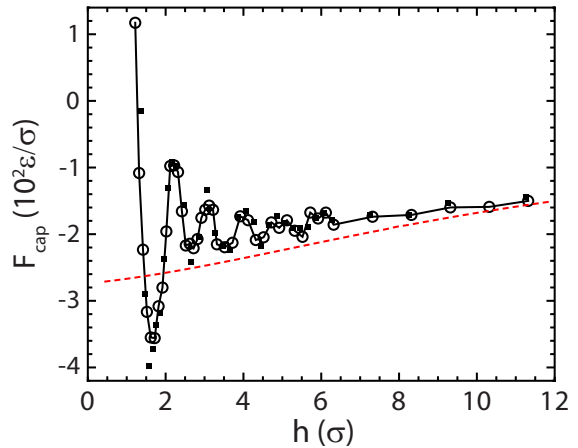}
\caption{(Color online) Force-separation curves for the incommensurate tip with $R=100\sigma$: 
elastic ($\bigcirc$) vs. rigid ($\blacksquare$) substrate.
The dashed line is the continuum solution.
The connecting lines are only guides for the eye.
}
\label{ElasticPlot}
\end{figure}

Figure~\ref{ElasticPlot} compares force-separation curves for the
incommensurate tip with elastic and rigid substrates.
In both cases the changes in $h$ reflect the displacement of the rigid tip
with respect to the substrate, which is the most accessible quantity 
in experiments.
For elastic substrates the actual change in separation between the closest tip and substrate atoms
may be slightly smaller due to compression of the substrate.
To remove this effect,
$h_0$ is calculated from the closest separation between tip
and substrate atoms before the tip contacts the drop.
It is then corrected to $h=h_0-2h_{ex}$ to include the excluded
volume effects discussed in the previous section.
The value of $h$ is then decreased by the change in tip height as the tip is lowered
into contact.

The results for elastic and rigid substrates in Fig.~\ref{ElasticPlot} are not
evaluated at exactly the same set of $h$ because of
the deformability of the elastic substrate, but are consistent within
statistical fluctuations for $h > 1\sigma$.
At smaller $h$ the dynamics in the liquid bridge become sluggish and large
viscous pressures can be generated as the separation is decreased.
Substrate elasticity couples to these transients, leading
to greater hysteresis between approach and retraction.
Since results in this regime are not reproducible, they are
excluded from Fig.~\ref{ElasticPlot} and subsequent plots,
but they may be relevant
to experiments.

To understand why substrate elasticity has little effect on capillary
adhesion in our simulations we estimate the magnitude of surface deformations using
continuum elastic results for a semi-infinite solid.
The Laplace pressure will produce a normal displacement of the
solid interface.
The effective modulus for normal substrate displacements
is $E^* \equiv E/(1-\nu^2)$ where $E$ is the Young's modulus and
$\nu$ is the Poisson's ratio \cite{johnson85}.
The normal strain is thus of order
$\Delta p / E^* = \gamma (1/r_1+1/r_2) /E^*$.
While we have chosen to use a relatively compliant solid,
the ratio $\gamma/\sigma E^* \sim 4 \times 10^{-3}$.
Since typical radii of curvature in our simulations are larger than $10\sigma$,
i.e., $\sigma/r \sim 0.1$,
the strains produced at the surface are only of order $10^{-4}$ \cite{footm3}.
Using results from contact mechanics \cite{johnson85},
one finds that the mean displacement of a semi-infinite
substrate from the constant pressure $\Delta p = \gamma (1/r_1+1/r_2) $
on a circle of radius $a$ is $16 (a/r_1 + a/r_2) \gamma /3 \pi E^*$.
In our simulations $a$, $|r_1|$, and $|r_2|$ are all of the same
order, so the predicted displacement is only of order $0.01\sigma$
and relatively constant across the liquid/solid interface.
It thus has little effect on force curves.
Deformation may be more important for small drops between
tips with macroscopic radii where $a \sim |r_2| \gg |r_1|$.

The meniscus also exerts a force in the region near the contact line.
The local pressure is of order $\gamma / \xi$,
where $\xi$ is the width of the liquid/vapor interface \cite{kirkwood49}.
For our fluids $\xi \sim 4\sigma$.
While the corresponding pressure is larger than the Laplace
pressure, the resulting strain is still only $\sim 10^{-3}$.
Since it enters over a narrow strip around the circumference,
its net effect on the substrate displacement is of the same order as the Laplace
contribution and negligible in our simulations.

Given the observed equivalence between simulations with rigid
and elastic walls, and the estimates of elastic effects from
the continuum theory, the remainder of the paper will focus on
results for rigid walls.
We have verified that all the trends and conclusions reached
are consistent with additional simulations with elastic walls
as long as $\gamma /E^* \ll \sigma$.
This condition is typically met for atomistic solids, but
the length $\gamma/E^*$ can be quite long for elastomers.
Their modulus is determined by the crosslink density and can
be reduced to extremely low values.
As shown in recent experiments and simulations,
surface tension dominates on lengths less than $\gamma/E^*$ and can lead
to large surface deformations \cite{dufresne11,style12, style13,dobrynin12}.
Elastomers only act like elastic solids above the crosslink spacing
which is typically several nanometers for systems with large $\gamma/E^*$.
The deviations from elasticity at smaller scales would affect capillary forces at the small surface separations considered here.

\subsection{Trends with Tip Radius and Liquid Bridge Volume}
\label{SecRad}

\begin{figure*}[htb]
\centering
\includegraphics[width=5in]{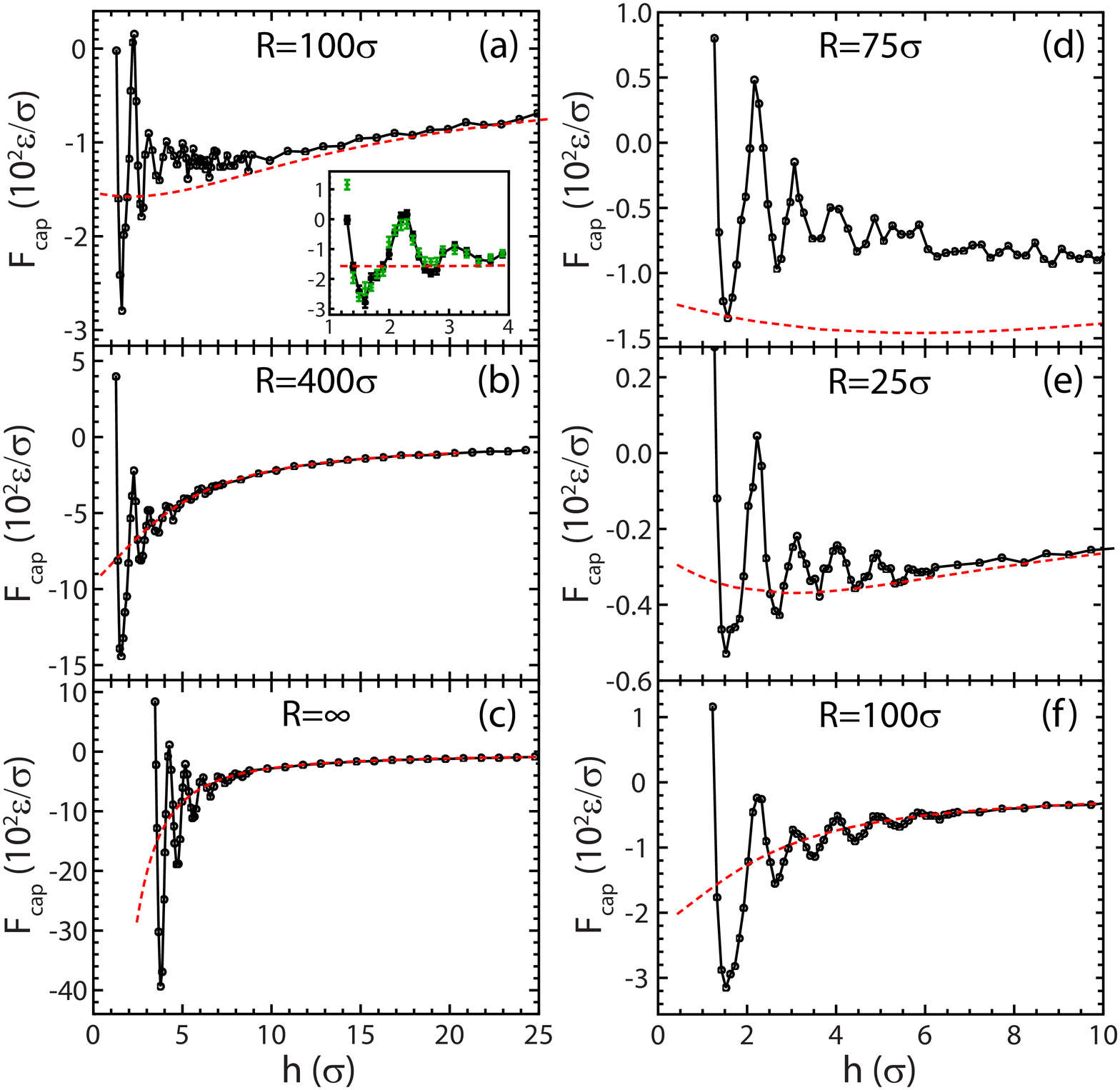}
\caption{(Color online) Force-separation curves for commensurate tips
with the indicated radii for (a-d) a large liquid
bridge with $V_l = 4.123\times 10^4 \sigma^3$ and
(e-f) a smaller bridge with $V_l = 2.390\times 10^3 \sigma^3$.
The dashed lines are the continuum solutions.
The connecting lines are only guides for the eye.
Symbols indicate the size of statistical uncertainties.
The inset of (a) shows that there is almost no hysteresis in the force between 
the approach ($\bigcirc$) and retraction ($+$).
}
\label{ForceRadPlot}
\end{figure*}

Figure~\ref{ForceRadPlot} shows force-separation curves for tips
with different radii and for liquid bridges with two different volumes $V_l$.
Commensurate tips are used in all cases to ensure that the changes
are not due to the difference in the tip structure.
We have also confirmed that the results represent steady state behavior.
In particular, as
illustrated in the inset of Fig. \ref{ForceRadPlot}(a),
data taken during the approach and retraction of the tip are consistent
within the statistical fluctuations.
There are much larger deviations from the continuum predictions 
(dashed lines in the main panels of Fig.~\ref{ForceRadPlot}).

In general we find that the continuum theory provides an accurate
description of $F_{cap}$ for separations bigger than some $h_c$ where
the subscript $c$ indicates {\it continuum}.
The length scale at which the percentage error exceeds our statistical
errors ($\sim 1-3$\%)
decreases as the tip radius increases at a fixed $V_l$.
For example for the larger liquid bridge with $V_l = 4.123\times 10^4 \sigma^3$,
$h_c$ drops from about 20$\sigma$ to less than 10$\sigma$
as $R$ increases from 100$\sigma$ (Fig. \ref{ForceRadPlot}(a)) 
to 400$\sigma$ (Fig. \ref{ForceRadPlot}(b)).
There is also a decrease in $h_c$ for the smaller bridge 
with $V_l= 2.390\times 10^3 \sigma^3$ as $R$ increases
from 25$\sigma$ (Fig. \ref{ForceRadPlot}(e)) to 100$\sigma$ (Fig. \ref{ForceRadPlot}(f)).
This trend is in part due to the rapid growth in the magnitude
of $F_{cap}$ with increasing $R$, which makes it harder to
see deviations as $R$ increases.
However the absolute error may also decrease as $R$ increases.
For example, for the larger liquid bridge the absolute error is $\sim 50 \epsilon/\sigma$
at $h=10\sigma$ for $R=75\sigma$ (Fig. \ref{ForceRadPlot}(d)),
but is reduced to $\sim 20\epsilon/\sigma$ for $R=100\sigma$ 
and $\sim 5\epsilon/\sigma$ for $R=400\sigma$.
Decreasing the volume of the liquid bridge at a fixed $R$ tends to reduce $h_c$, as well
as all other dimensions of the drop.

The deviations from continuum theory at $h < h_c$ have two forms.
Near $h_c$ there tends to be a systematic upward shift in the data
that indicates a less adhesive force.
There are also oscillations for $h < h_o$ that grow
rapidly in magnitude as $h$ decreases.
The value of $h_o$ and the magnitude of oscillations tend
to be bigger for larger $R$.
The oscillations also become more pronounced and coherent as $V_l$
is reduced.

Force oscillations result from the layering of liquid molecules in the gap
between the tip and substrate.
This layering was observed in early simulations
of flat surfaces \cite{toxvaerd81}
and the associated force oscillations were first measured between
the nearly parallel plates of an SFA \cite{horn81}.
Recent experiments have found force oscillations for
AFM tips with radii of curvature similar to the tips
studied here \cite{jarvis00}.
Layering leads to attractive forces when the separation is slightly
larger than an integer multiple of the equilibrium layer spacing,
and repulsion when the separation is reduced below this 
optimal spacing.
The alternating of these two regimes produces an oscillatory
signature in the force-separation curve.

\begin{figure}[htb]
\centering
\includegraphics[width=3in]{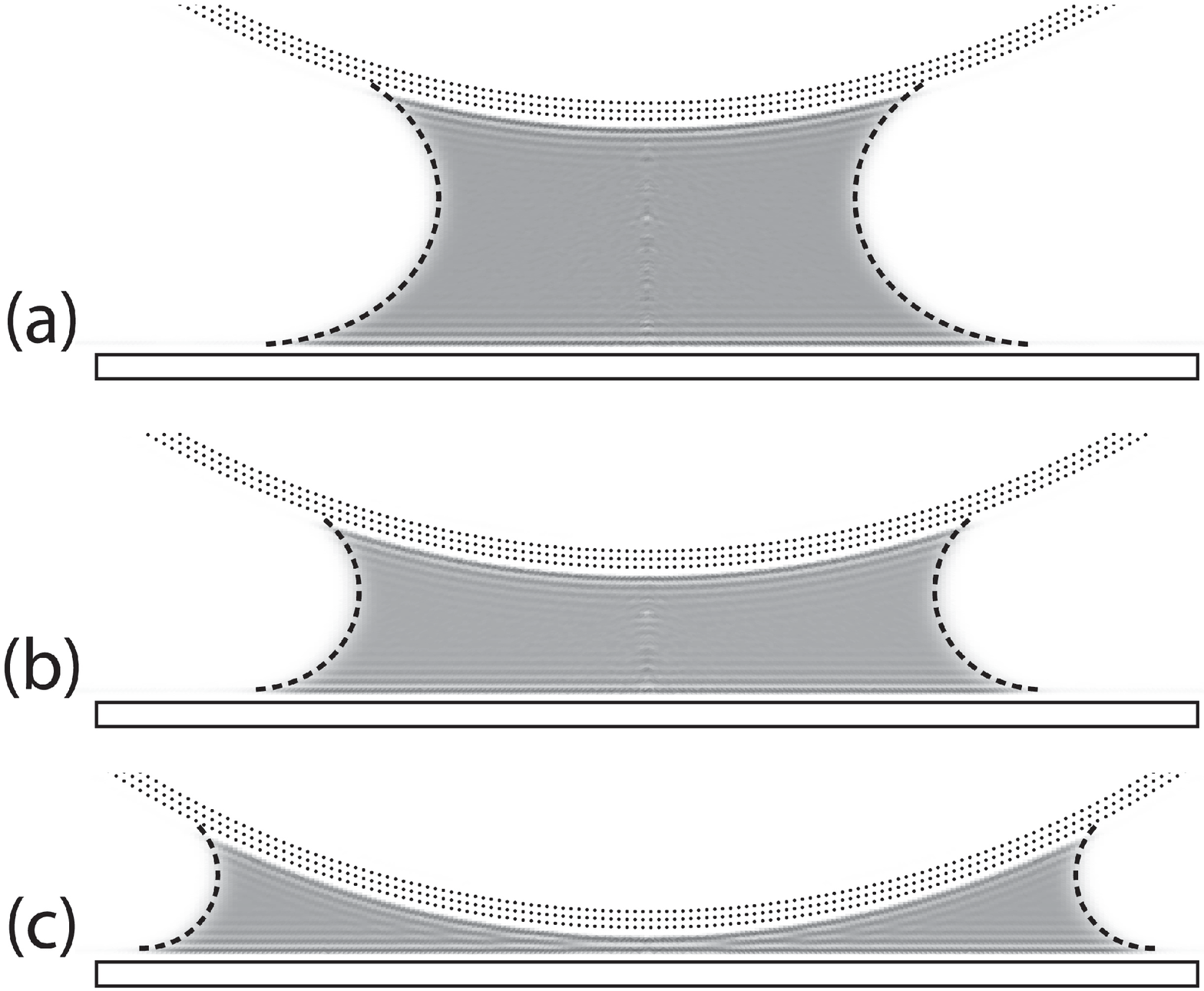}
\caption{Layering of the liquid between the tip and substrate
for the bent commensurate tip
with $R=100\sigma$ and $\theta =12^\circ$ at (a) $h=21.9\sigma$,
(b) $h=11.9 \sigma$, and (c) $h=1.9 \sigma$.
The gray scale
shows the density of liquid as a function of height above the substrate
and radial distance from the central axis of the tip.
Results are averaged over angles and there is more noise at the
tip center because a smaller volume contributes to the average.
Dots indicate the positions of tip atoms in a slice through
its central axis and dashed lines represent the continuum predictions for
the interface profile  (Sec. ~\ref{continuum_theory}).
}
\label{layering}
\end{figure}

Figure \ref{layering} shows the layered structure
of the liquid in the gap between the tip and substrate.
The strongest density peaks correspond to layers spaced by $\sim \sigma$
from either the flat wall or the curved tip.
Layering continues into the bulk, but with a decreasing amplitude
as the distance from the wall increases.
Previous studies of flat surfaces have shown that
the range of oscillations is comparable
to that of the pair correlation function of the liquid, and that
the magnitude
of oscillations increases with the magnitude and steepness of the
potential from the solid
\cite{thompson90a,magda85,toxvaerd81}.
When $h$ is larger than the range of oscillations the central
region of the drop behaves like a continuous liquid (Fig.~\ref{layering}(a)).
At smaller $h$, the layering from opposing surfaces begins
to interfere, resulting in the oscillatory forces.
A small amount of layering is present throughout the gap
at $h=11.9\sigma$ (Fig.~\ref{layering}(b)) and the layering is pronounced
at $h=1.9\sigma$ (Fig.~\ref{layering}(c)).

The effect of liquid layering on $F_{cap}$ is largest
for flat surfaces where the separation is constant across the gap.
The curvature of the tip in Fig. \ref{layering} leads to variation
in the separation between solid surfaces and thus in the number of layers.
As a result, the layering force varies in sign and magnitude with
$\rho$, the radial distance from the central axis of the tip, 
reducing the net contribution to $F_{cap}$.
These observations explain the trends in Fig. \ref{ForceRadPlot}.
The oscillations in $F_{cap}$ decrease with decreasing $R$ because
the more rapid changes in separation at small $R$ lead to cancelation of the layering force
from different parts of the tip.
Decreasing $V_l$ reduces the range of separations filled
by the liquid and thus enhances the force oscillations.
The local layering forces are discussed
further in Sec.~\ref{force_components}.

\subsection{Effects of Atomic Structure of Tip and Contact Angle}

The atomic structure of the tip affects both the macroscopic contact
angle and the degree of liquid layering in the gap.
Figures \ref{ElasticPlot} and \ref{ForceRadPlot}(a) are for the
bent incommensurate and commensurate tips with $R=100\sigma$, respectively.
The incommensurate tip has a higher density, producing a more attractive
potential and a smaller contact angle.
As a result, both the continuum solution and simulation results show
a more attractive capillary force for the incommensurate tip.
The magnitude of oscillations is similar until $h < 4 \sigma$, where
it rises more rapidly for the commensurate tip.
This difference reflects the stronger in-plane order induced by two commensurate surfaces.
Fluid monomers tend to order epitaxially with atoms centered over
favorable sites along the substrate.
When the epitaxial order induced by both walls adds coherently,
as in the gap between two atomic lattices commensurate to each other,
the force oscillations become stronger and a crystalline order may even
be induced in the fluid \cite{thompson90a}.

\begin{figure}[htb]
\centering
\includegraphics[width=3in]{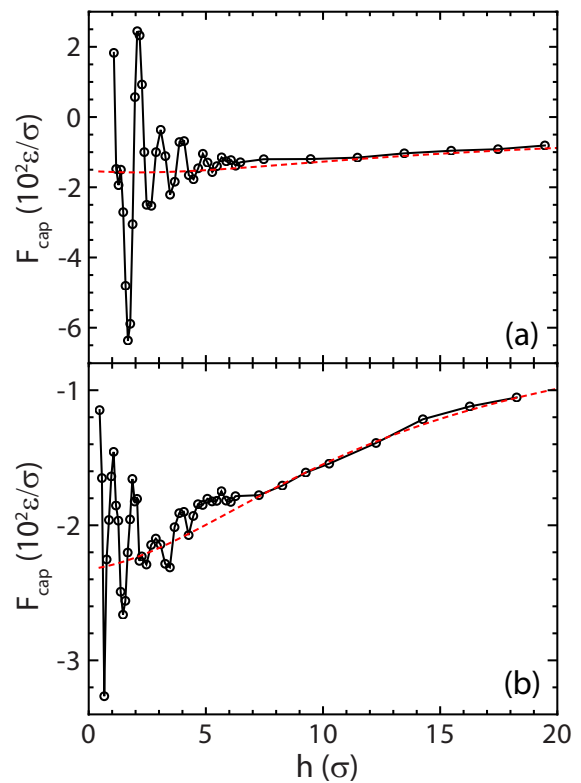}
\caption{(Color online) Force-separation curves for 
(a) the stepped tip and (b) the amorphous tip with $R=100\sigma$
and the large liquid bridge with $V_l = 4.123\times 10^4 \sigma^3$.
Note that the magnitude of oscillations is about three times smaller for the amorphous tip.}
\label{ForceStepAmorPlot}
\end{figure}

Figure \ref{ForceStepAmorPlot} shows results for the stepped and
amorphous tips with $R=100\sigma$ and the large liquid bridge with 
$V_l = 4.123\times 10^4 \sigma^3$.
Results for the stepped tip lie close to the continuum solution for the same
contact angle ($\theta_t=75^\circ$) as the curved commensurate tip.
The smaller contact angle ($\theta_t=66.3^\circ$) for the amorphous tip
leads to larger adhesion,
though smaller than for the incommensurate tip ($\theta_t=61.6^\circ$).

\begin{figure}[htb]
\centering
\includegraphics[width=3in]{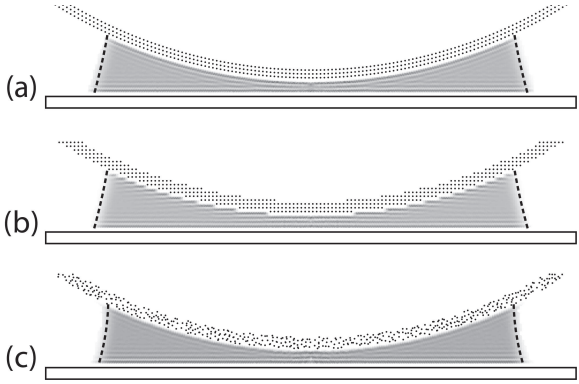}
\caption{Layering of liquid molecules in the gap between the substrate and various tips 
with $R=100\sigma$ and $V_l = 4.123\times 10^4 \sigma^3$:
(a) the commensurate tip at $h=2.30 \sigma$; 
(b) the stepped tip at $h=2.27 \sigma$;
and (c) the amorphous tip at $h=2.26 \sigma$.
The gray scale
shows the density of liquid as a function of height above the substrate
and radial distance $\rho$ from the central axis of the tip.
Results are averaged over angles and there is more noise at the
tip center because a smaller volume contributes to the average.
Dots indicate the positions of tip atoms in a slice through
its central axis and dashed lines represent the continuum predictions for
the interface profile (Sec. ~\ref{continuum_theory}).
}
\label{layering2}
\end{figure}

Changes in the atomic structure of the tip lead to more pronounced variations
in the oscillatory forces from layering.
While the stepped and bent commensurate tips have the same contact
angle, the stepped tip gives much stronger oscillatory forces.
The reason is evident from Fig. \ref{layering2}.
The surface of the stepped tip consists of
terraces with fixed separation from the substrate, and the
terrace spacing is close to the layer spacing.
Thus the layering forces tend to add coherently over the tip
surface rather than oscillate as a function of $\rho$ and partially cancel each other.
Note that for the stepped tip the size of terraces is not unique for a given tip
radius.
For the case shown here, the first terrace has radius 8.9$\sigma$,
but smaller and larger values can be obtained by varying the position of
the center of the sphere used to cut the solid into a tip.
These variations will change the magnitude of force oscillations and
there should be similar variations between crystalline AFM tips
with the same nominal radius.

In contrast to the stepped tip, the random roughness on the amorphous surface
suppresses force oscillations.
Even though height variations are less than one atomic diameter,
they are sufficient to change the magnitude and sign of the layering force.
The force oscillations nearly cancel except for tip separations 
that accommodate less than 2 or 3 liquid layers (i.e., $h<3\sigma$).

\begin{figure}[htb]
\centering
\includegraphics[width=3in]{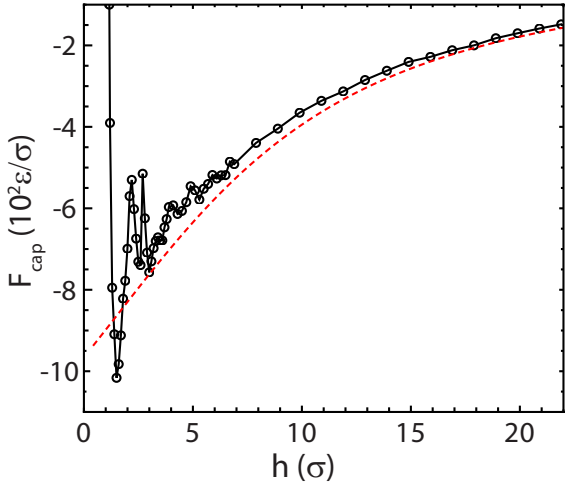}
\caption{(Color online) The force-separation curve for the commensurate tip with
$R=100\sigma$, $V_l = 4.123\times 10^4 \sigma^3$,
and contact angles $\theta_t=\theta_s=12^\circ$.}
\label{ForceSCAPlot}
\end{figure}

The changes in contact angle with tip structure in Fig. \ref{ForceStepAmorPlot}
are relatively small (Table \ref{table_angle}) because $\epsilon_{sf}$ is held fixed.
Figure \ref{ForceSCAPlot} shows $F_{cap}$ vs. $h$ for the commensurate tip with $R=100\sigma$ when $\theta_s$
and $\theta_t$ are both reduced to $12^\circ$ by increasing
$\epsilon_{fs}$ from $0.8\epsilon$ to $1.08\epsilon$.
Both the continuum prediction and the MD results for
$F_{cap}$ increase in magnitude by a factor of about 4.
This increase can be understood as follows.
Comparing Fig.~\ref{layering2} to Fig.~\ref{layering}, it is clear
that the radius of curvature $r_1$ has a much smaller magnitude
for $\theta=\theta_t=\theta_s=12^\circ$.
For $\theta<\pi/2$, the radii of curvature $r_1$ and $r_2$ have opposite signs with $r_1<0$ and $r_2>0$.
At a large $\theta$ in this range (e.g., the case with $\theta=75^\circ$), $|r_1|$ is large and the magnitude of $\Delta p=\gamma (1/r_1+1/r_2)$ is small and may even be positive,
giving a repulsive contribution to $F_{cap}$ in Eq. \ref{fcap} and counteracting the surface tension term.
The net result is reduced capillary adhesion.
For $\theta=12^\circ$, $-r_1 \ll r_2$ and the large
negative $\Delta p$ gives a strong attractive contribution to
$F_{cap}$ that enhances the attractive meniscus force and leads to large capillary adhesion.

The deviations of the MD results from the continuum theory in
Fig. \ref{ForceSCAPlot} for $\theta=12^\circ$ are comparable to those in other
figures shown previously for $\theta=75^\circ$.
Once again the capillary force shows oscillations at $h < h_o \sim 5\sigma$ and is
less attractive for $h < h_c \sim 15\sigma$.
We now consider the origins of these differences quantitatively.

\subsection{Components of Capillary Force}
\label{force_components}

As introduced in Sec.~\ref{continuum_theory}, 
the continuum theory expresses $F_{cap}$ in terms of meniscus and Laplace
pressure terms.
Both depend on the shape of the interface through the radius of
the contact line $a_i$ and the interface curvature described by $r_i$.
Figures \ref{layering} and \ref{layering2}
include comparisons of the interface shape from
simulations with the continuum predictions (dashed lines).
Within the intrinsic uncertainty due to the interface halfwidth
($\sim 1 - 2 \sigma$), the continuum theory is in excellent agreement
with the simulations.
Thus the radii of curvature $r_i$ that determine the capillary pressure
in the continuum theory are consistent with the simulation results.

\begin{figure}[htb]
\centering
\includegraphics[width=3in]{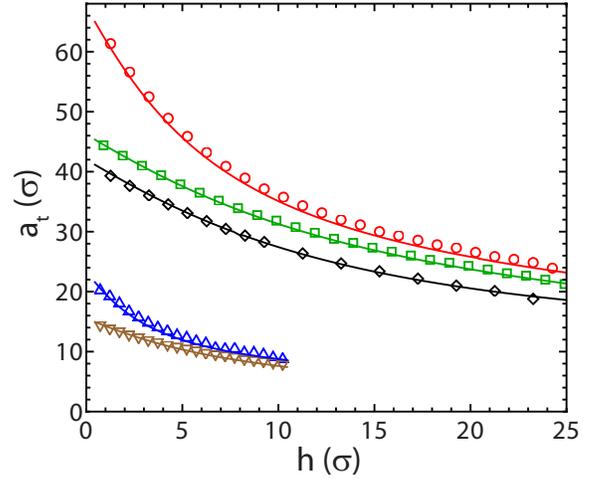}
\caption{(Color online) The radius of the contact line on the commensurate
tip for some of the cases in Fig. \ref{ForceRadPlot}.
The contact angles are $\theta_t=\theta_s=75^\circ$.
The top three data sets are for the large liquid bridge ($V_l = 4.123\times 10^4 \sigma^3$) and tips with $R=75\sigma$ (black diamond), $R=100\sigma$ (green square) and $R=400\sigma$ (red circle).
The bottom two data sets are for
the small liquid bridge ($V_l = 2.390\times 10^3 \sigma^3$) and tips with $R=25\sigma$ (brown downward triangle) and $R=100\sigma$ (blue upward triangle).
Solid lines show the continuum predictions.
}
\label{atip}
\end{figure}

Figure \ref{atip} shows that the simulation results for $a_t$ (the radius of contact circle on the tip surface) also agree with the continuum theory.
The radius $a_t$ is computed from either the first moment of the probability that a tip
atom interacts with the fluid or by the ratio between the first and second moments.
It is then corrected for the interaction range.
The difference in results from these analysis methods is less than 0.5 $\sigma$, which is comparable to the largest deviations from the continuum theory.
The finite width of the interface appears to lead to a very small outward
shift in the radius $a_t$, but this shift is too small to explain the observed
deviations from the continuum predictions of the capillary force.

\begin{figure*}
\centering
\includegraphics[width=6in]{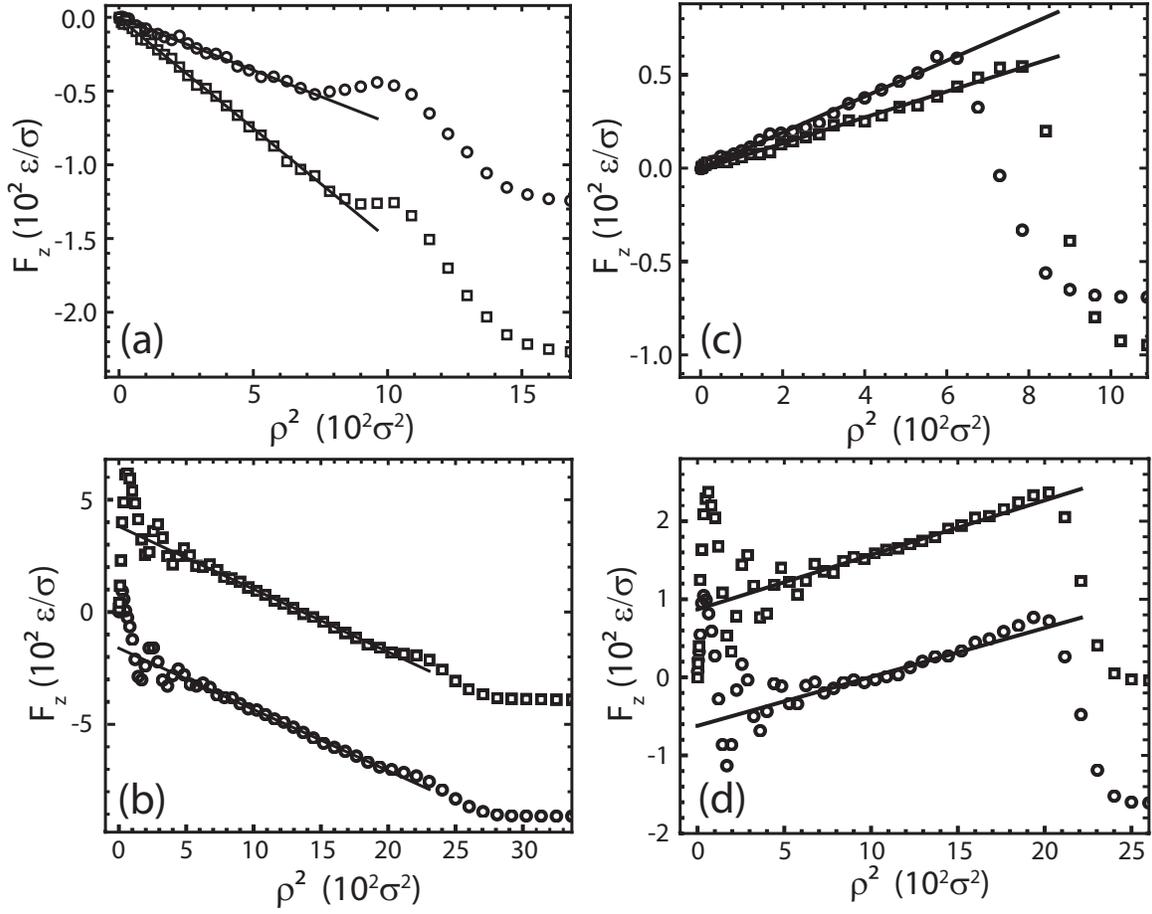}
\caption
{Integral of the vertical force $F_z(\rho)$ within a circle of radius $\rho$ 
vs. $\rho^2$ for the commensurate tip with $R=100\sigma$ and 
the large liquid bridge with $V_l = 4.123\times 10^4 \sigma^3$.
The left panels are for $\theta_t=\theta_s=12^\circ$
at various tip-surface separations:
(a) $h=24.9\sigma~(\bigcirc)$ and $15.9\sigma~(\Box)$;
(b) $h=1.4\sigma~(\bigcirc)$ and $1.3\sigma~(\Box)$.
The right panels are for $\theta_t=\theta_s=75^\circ$
at various tip-surface separations: 
(c) $h=24.9\sigma~(\bigcirc)$ and $15.9\sigma~(\Box)$;
(d) $h=1.4\sigma~(\bigcirc)$ and $1.2\sigma~(\Box)$. 
The linear fits give a slope $\pi p_{z}$ that varies slowly with $h$
and an offset $F_{in}$ that oscillates rapidly with $h$.}
\label{ForAreaPlot}
\end{figure*}

Since the interface shape, including $a_t$ and $a_s$, 
is consistent with the continuum theory, there must be 
other factors that lead to the observed shift in the total capillary force from the continuum prediction. 
These could be deviations in the Laplace pressure $\Delta p$ 
or the liquid/vapor interfacial tension $\gamma$, 
or other contributions not captured in continuum theory.
In Sec.~\ref{SecRad} we have discussed force oscillations at small $h$ and
related them to layering of the liquid in the gap between the tip and substrate. 
Here we make this analysis more quantitative. 
Since the local separation increases 
continuously with lateral distance $\rho$ from the center of the gap, 
the layering of liquid also 
varies with $\rho$, as does the local force between the liquid bridge and wall.
Our method for analyzing the net effect of oscillations 
is illustrated for the commensurate tip with $R=100\sigma$ in Fig.~\ref{ForAreaPlot}.
Results are shown for contact angles $\theta_t=\theta_s=75^\circ$
and $\theta_t=\theta_s=12^\circ$, which give different signs for the Laplace pressure.

In Fig.~\ref{ForAreaPlot}, rather than plotting the local force directly, 
the integral of the force $F_z(\rho)$ on the substrate within a circle of radius $\rho$ 
around the central axis of the tip is shown 
as a function of $\rho^2$.
Using the integrated force
reduces noise and separates different contributions to the calculated capillary force.
At the outer edge of the contact there is 
a strong attractive contribution from the liquid/vapor interfacial tension.
At smaller $\rho$, the slope of the curve gives the local pressure acting on
the substrate, $p_{z} = \frac{1}{\pi}\frac{\partial F}{\partial \rho^2}$.
When the tip-substrate separation $h$ is large, 
$F_z(\rho)$ grows linearly with area $\pi\rho^2$ until the liquid/vapor interface is reached, 
indicating a constant vertical pressure $p_{z}$ between the liquid bridge and wall. 
When $h$ gets smaller than $h_c$, oscillations of $F_z(\rho)$ with $\rho^2$ 
are obvious in the central region of the gap where $\rho$ is small.
At larger $\rho$ there is a constant normal pressure again 
and $F_z(\rho)$ grows linearly with area.
For the more wetting walls ($\theta_t=\theta_s=12^\circ$),
$F_z(\rho)$ decreases with $\rho^2$,
indicating a negative Laplace pressure and
an attractive contribution to $F_{cap}$.
For the same tip and $V_l$
but with $\theta_t=\theta_s=75^\circ$
the Laplace pressure is
positive, i.e., repulsive.

Linear fits in Fig.~\ref{ForAreaPlot} allow us 
to extract a slope corresponding to a mean vertical pressure $p_{z}$ 
at intermediate $\rho$ and
an offset $F_{in}$ associated with the total 
contribution of the layering forces in the central region.
As shown in Fig. \ref{ForAreaPlot},
$p_{z}$ and its contribution to the capillary force, $F_p = p_{z} \times \pi a_t^2$, 
vary slowly with $h$, as expected from the continuum theory.
However, $F_{in}$ oscillates rapidly at small $h$ and 
it is this term that makes up the oscillatory part of $F_{cap}$.

\begin{figure*}[htb]
\centering
\includegraphics[width=6in]{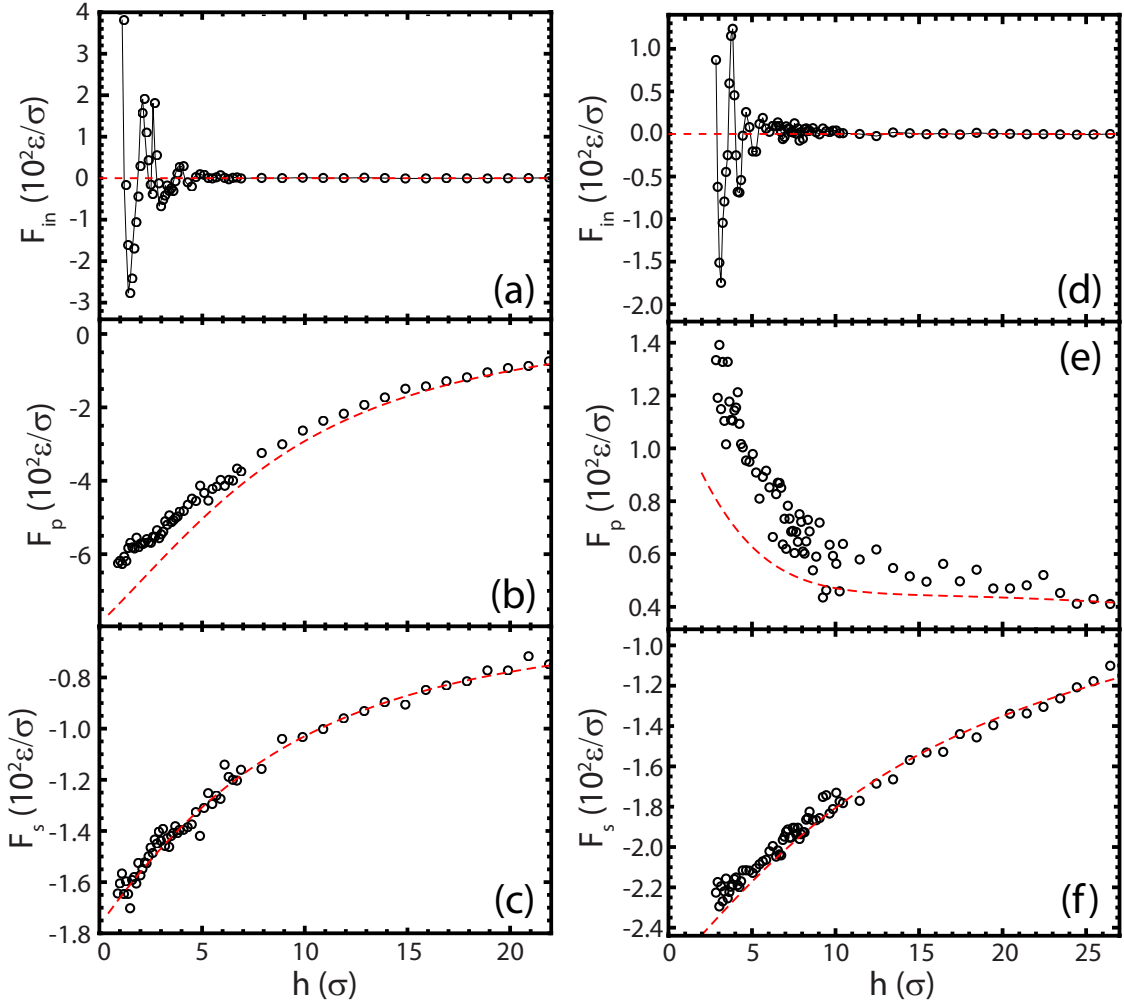}
\caption[Three components of capillary force]
{(Color online) Three components of $F_{cap}$ for the commensurate tip with 
$R=100\sigma$, the large liquid bridge with $V_l = 4.123\times 10^4 \sigma^3$, 
and (a-c) contact angles $\theta_t=\theta_s=12^\circ$ or
(d-f) contact angles $\theta_t=\theta_s=75^\circ$.
Panels
(a, d) show the oscillatory component $F_{in}$ from the layering force,
(b, e) show the Laplace pressure component $F_p$, and
(c, f) show the liquid/vapor interfacial tension component $F_s$.}
\label{ForThreePlot}
\end{figure*}

Using the fits illustrated in Fig.~\ref{ForAreaPlot}
three contributions to the capillary force can
be identified,
\begin{equation}
F_{cap}=F_{in} + F_p + F_s \ \ ,
\end{equation}
where $F_s=2\pi a_t \gamma \sin\phi_t$ is the contribution from the
interfacial tension around the circumference of the meniscus.
Since the continuum theory provides an excellent description of the interface
shape, it gives accurate values of $\phi_t$ as well as $a_t$ (Fig. \ref{atip}).
Values of $F_s$ obtained by using either the fit or predicted values of $\phi_t$ and
$a_t$ and from $F_s=F_{cap}-F_{in}-F_p$ are equivalent within statistical
errors.
The latter is used in plotting the three separate components of $F_{cap}$
in Fig. \ref{ForThreePlot}.
Consistent values of $F_p$ were also obtained from the fit $p_z$ with either directly measured
$a_t$ or the value of $a_t$ calculated with the continuum theory.

Figure~\ref{ForThreePlot} shows the three components of $F_{cap}$ for
two contact angles, $\theta_t=\theta_s=12^\circ$ and $75^\circ$.
The surface contribution
$F_s$ is consistent with the continuum prediction
for $h$ down to $0.5\sigma$ for all cases considered.
The underlying reasons are that the interface shape is correctly described by continuum theory and the interfacial tension $\gamma$ of the liquid/vapor
interface remains near the bulk value.
The latter was confirmed by independent direct calculations of $\gamma$
as a function of curvature and distance from a solid wall.
It is also understood the fluid film is thickest at the circumference of the meniscus,
which limits atomistic effects at the interface.

The deviations from continuum theory are all associated with pressure deviations
that contribute to $F_{in}$ and $F_p$.
$F_{in}$ is only significant for the range of $h < h_o \approx 6\sigma$
where the total force shows oscillations.
The oscillations have a larger magnitude for the more wetting case that also
has stronger wall/fluid interactions.
They represent variations in the ease of packing discrete atoms
between the confining walls and are not included in the continuum theory.

The contribution from the Laplace pressure in outer regions of the drop, $F_p$,
shows a systematic deviation from the continuum theory for $h < h_c \approx 18\sigma$.
The observed force is always less attractive than the continuum predictions.
The magnitude of the shift is consistent with the
systematic deviations in the
total forces shown in Fig.~\ref{ForceRadPlot}(a) and Fig.~\ref{ForceSCAPlot}.
For the less wetting case and the commensurate tip with $R=100\sigma$ the average adhesion is reduced by as much as a factor of 2 
[Fig.~\ref{ForceRadPlot}(a)]. 
The reduction is more pronounced for even smaller tips [Fig.~\ref{ForceRadPlot}(d)].

Given that the continuum theory gives the correct interface shape
and the surface tension retains the bulk value,
the deviation of $F_p$ from the continuum prediction implies
a failure of the Laplace equation (Eq. \ref{Laplace})
for the pressure in the outer region of the drop.
Note that pressure is a tensor and
only the vertical component, $p_{zz}$, contributes to $F_p$ and
$F_z(\rho)$ in Fig.~\ref{ForAreaPlot}.
If the interior of the liquid bridge was isotropic, then 
the pressure tensor would be hydrostatic: $p_{zz}= p_{xx}=p_{yy}$.
However, the confining solids can introduce anisotropy.
In the case of thin liquid films, there is commonly a disjoining pressure
normal to the film
that acts to increase or decrease the thickness \cite{degennes85}.
As we now show, the in-plane components of the pressure are quite
close to the values predicted by the Laplace equation,
but the out of plane component that determines $F_{cap}$ is
consistently more positive.

\begin{figure}[htb]
\centering
\includegraphics[width=3in]{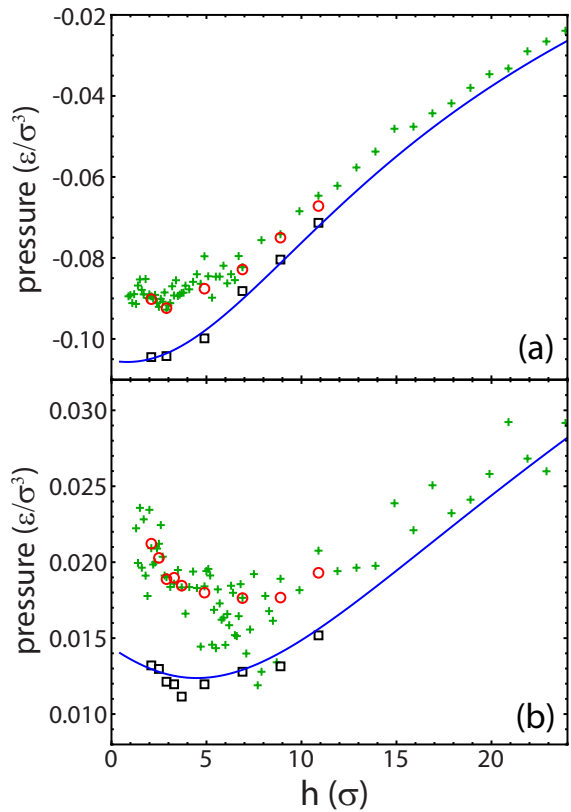}
\caption[Pressure anisotropy in a liquid bridge]
{(Color online) Components of pressure tensor for the commensurate tip with 
$R=100\sigma$, $V_l = 4.123\times 10^4 \sigma^3$, 
and contact angles (a) $\theta_t=\theta_s=12^\circ$
and (b) $\theta_t=\theta_s=75^\circ$.
Symbols show
the out-of-plane component $P_{zz}$ (red $\bigcirc$)
and the in-plane component $P_{rr}$ (black $\square$)
of the pressure tensor in the fluid and the pressure on the substrate
$p_z$ (green $+$) from the fits shown in Fig.~\ref{ForAreaPlot}.
Blue solid lines show the
continuum prediction for the Laplace pressure $\Delta p$
from Eq.~\ref{Laplace}.}
\label{StressPlot}
\end{figure}

We computed the pressure tensor in the liquid bridge 
directly using an algorithm developed by Denniston et al. \cite{denniston04},
which is based on an expression for the pressure tensor derived 
by Irving and Kirkwood \cite{kirkwood50}.
The out-of-plane component $P_{zz}$ and the in-plane component $P_{rr}$ are
averaged over values of $r$ that are large enough to avoid pressure
oscillations and small enough to be inside the meniscus (Fig. \ref{ForAreaPlot}).
Figure~\ref{StressPlot} compares the variation in these components of the
pressure tensor with the prediction for the Laplace pressure $\Delta p$
from the continuum theory (line) and the
measured pressure on the solid substrate, $p_z$.
For both contact angles, $P_{rr}$ is close to the predicted Laplace pressure at
all separations.
As expected from force balance considerations, $P_{zz}$ is also
close to the normal force exerted on the substrate.
The pressure anisotropy $P_{zz} - P_{rr}$ increases as $h$ decreases
in exactly the same way that the pressure contribution to $F_{cap}$
deviates from the continuum prediction.
Similar results were observed with various tips and liquid bridge volumes.
As found above, the pressure anisotropy always leads to more repulsive interactions and grows in magnitude as $h$ decreases below about 20$\sigma$.
For $h<5\sigma$ it represents a substantial correction to the Laplace pressure
contribution to $F_{cap}$.
We conclude that the systematic deviation in $F_{cap}$ from the continuum theory results
from the pressure anisotropy in the liquid bridge.
The physical origin is similar to that giving the disjoining pressure of thin liquid films.

\section{Summary and Conclusions}

In this paper, we used MD simulations to study the capillary forces induced by a liquid meniscus 
that connects a rigid spherical tip and a flat substrate.
We showed that the substrate elasticity has a negligible effect on 
the total capillary adhesion force, $F_{cap}$,
for most substrates, although new effects may arise for very soft elastomers
\cite{dufresne11,style12, style13,dobrynin12}.
To reduce the computational burden, the substrate is treated as rigid in most simulations reported in this paper.

The value of $F_{cap}$ was calculated as a function of 
the tip-surface separation $h$ for a range of tip geometries. 
We showed that if $F_{cap}$ is calculated when the tip is displaced at a constant velocity $v$, 
a hysteresis with magnitude depending on $v$ occurs between approach and retraction.
To measure equilibrium capillary forces, the separation $h$ is gradually varied and 
the liquid bridge is allowed to relax to its equilibrium state at a given $h$ before 
$F_{cap}$ is computed.
As expected, the force-separation curves obtained in this way do not exhibit hysteresis at most separations.
However, films that are only 1 to 2 layers thick may enter a glassy state
that resists equilibration \cite{gee90,thompson92,granick91,cheng10pre,thompson92}.
This immobile layer may affect experimental
determinations of the onset of tip substrate contact. 

Our results show that for a typical $F_{cap}$ vs. $h$ curve, 
there exist two critical separations $h_c$ and $h_o$.
The separation $h_c$ is typically about $20\sigma$ ($\sim 10$nm) but can vary with 
the tip radius $R$ and the liquid bridge volume $V_l$.
The separation $h_o$ is roughly $10\sigma$ ($\sim 5$nm).
When the separation $h$ is larger than $h_c$, 
the magnitude of $F_{cap}$ grows as $h$ decreases and 
agrees with the continuum prediction. 
This agreement shows that the continuum theory of capillaries works down to surprisingly small scales.
When $h$ is smaller than $h_o$, 
the adhesion force $F_{cap}$ oscillates strongly with $h$ because of molecular layering in the liquid bridge.
At an intermediate separation $h_o < h < h_c$, the adhesion $F_{cap}$
is more positive (less attractive) compared with the continuum prediction. 
The average adhesion can be reduced from 
the continuum prediction by a factor of $2$ or even change sign.

The magnitude of $F_{cap}$ grows as the tip radius $R$ increases because 
the Laplace pressure term becomes more attractive 
(or less repulsive when it is positive) and adds to the 
meniscus contribution [Eq.~\ref{fcap}]. 
The oscillations in $F_{cap}$ at small $h$ decrease 
with decreasing $R$ or increasing liquid bridge volume $V_l$.
The reason is that the changes in local separation between the
tip and substrate lead to cancellations of the layering force that 
is responsible for the oscillations. 
This cancellation is more pronounced for smaller tips or larger liquid bridges.

For tips with the same $R$, the atomic geometry affects $F_{cap}$ in two ways.
One is by changing the contact angle $\theta_t$. 
The atomic structure also affects the layering of liquid in the gap between 
the tip and substrate and thus the layering force contribution 
to $F_{cap}$. In general, oscillations in $F_{cap}$ at small $h$ 
for atomically smooth tips (e.g., the commensurate or incommensurate tips) 
are more coherent and stronger 
than those for tips with rougher surfaces (e.g., the amorphous tip). 
However, the stepped tip leads to even stronger oscillations because the changes in 
local separation are slower (each terrace is actually flat) than for the corresponding commensurate tip.
For the commensurate tip with $R=100\sigma$, $F_{cap}$ becomes much more 
attractive when the contact angles $\theta_t$ and $\theta_s$ are reduced 
from $75^\circ$ to $12^\circ$. The adhesion becomes stronger because 
the Laplace pressure contribution to $F_{cap}$ changes from
repulsion to attraction, which more than compensates for the reduction of
the attractive meniscus contribution.

Direct calculations of the density profile of the liquid bridge indicate that 
the interface shape is consistent with the continuum theory. 
The radii of contact circles ($a_t$ and $a_s$) on the tip and substrate surfaces agree
with the continuum predictions within an intrinsic uncertainty 
associated with the half-width of the liquid/vapor interface. 
Such calculations also show clearly the layering of liquid near the solid walls 
and the interference of the layering from two walls at small $h$ 
(Figs. \ref{layering} and \ref{layering2}.

Our simulations show that generally $F_{cap}$ contains three contributions: 
an oscillatory part that originates from the molecular layering force 
in the inner region of the liquid bridge ($F_{in}$); 
a contribution from the liquid/vapor interfacial tension ($F_s$); 
and a contribution ($F_p$) from the average normal pressure $p_{zz}$.
The inner term $F_{in}$ oscillates between positive and negative values as $h$ varies and 
is only significant when $h<h_o$. 
This term represents the change in the free energy of the liquid bridge when $h$ is varied.
It is not captured in any continuum models developed so far but 
has been observed in SFA and AFM experiments
\cite{horn81,thompson90a,jarvis00}.
The meniscus contribution $F_s$ agrees with the continuum predictions for $h$ down to 
the order of $2\sigma$ ($\sim 1$nm). This is consistent with the facts that the liquid/vapor interface shape
is accurately described by the continuum theory and the interfacial tension $\gamma$
does not change appreciably when the interface approaches the solid wall.
The mean normal pressure term $F_p$ agrees with the continuum predictions for $h>h_c$, 
but becomes consistently more positive (less attractive or more repulsive) 
when $h$ is reduced below $h_c$.
This shift is found to be due to the anisotropy of the pressure tensor 
in the liquid bridge. 
The in-plane component $p_{rr}$ is consistent with the Laplace pressure $\Delta p$ predicted by the continuum theory. 
However, it is the out-of-plane component $p_{zz}$ that determines the
normal forces on the tip and substrate and thus the
pressure contribution to $F_{cap}$.
This term is always more positive than the predicted $\Delta p$.

In this paper, we focus on $F_{cap}$ as a function of $h$. 
As many experiments indicate, capillary forces also contribute significantly to sliding friction 
and lead to unusual velocity dependence (dynamics) and time dependence (kinetics) \cite{binggeli94,ohnishi02,riedo02,riedo05,noel12}. 
It would be interesting to extend the current study to the cases where the tip
is displaced laterally so that the liquid bridge is dragged over the substrate surface. 
The normal force (adhesion) and lateral force (friction) can be calculated simultaneously
and their correlations and variations can be studied. 
This would help us understand the behavior of capillary forces when surfaces are in relative motion.
One example is the stability of granular materials when they are 
sheared or vibrated.

\section*{Acknowledgments}
This material is based upon work supported by the National Science Foundation
under Grant No. CMMI-0709187, DMR-0454947, DMR-1006805 and the Air Force Office of Scientific
Research under Grant No.~FA9550-0910232.


\begin{thebibliography}{101}
\bibitem[{foo({\natexlab{a}})}]{footm2}
\bibinfo{note}{Any direct interactions between the two solids can be described
  separately.}
\bibitem[{foo({\natexlab{b}})}]{footm3}
\bibinfo{note}{Note that some macroscopic systems can yield through dislocation
  motion even at such small strains. However, the radii of curvature tend to be
  larger at these scales and the threshold for plastic flow is higher at
  submicron scales.}
\end{thebibliography}
\end{document}